EXACT LAG SYNCHRONIZATION IN TIME DELAY SYSTEMS AND WHY EXPERIMENTAL LAG TIMES CAN DIFFER FROM THEORETICAL PREDICTIONS.

E. M. Shahverdiev [1], S. Sivaprakasam and K. A. Shore [2]

School of Informatics, University of Wales, Bangor, Dean Street, Bangor, LL57 1UT, Wales, UK

ABSTRACT

We present analytical investigation of *exact* lag synchronization between two unidirectionally coupled identical time delay systems with two characteristic delay times, where the delay time in the coupling is different from the delay time in the coupled systems themselves. We show that the lag time in synchronization of master and slave systems is the difference between the coupling delay time and the time delay in the coupled systems themselves. Also, for the first time we demonstrate that parameter mismatches can explain the experimental observation that the lag time is equal to the coupling delay.

Seminal papers on chaos synchronization [1] have stimulated a wide range of research activity [2]. Chaos synchronization in coupled systems have been especially extensively studied in the context of laser dynamics, electronic circuits, chemical and biological systems [2]. Chaos synchronization can be applied in secure communications, optimization of nonlinear system performance, modeling brain activity and pattern recognition phenomena [2]. Due to finite signal transmission times, switching speeds and memory effects time delay systems are ubiquitous in nature, technology and society [3]. Therefore the study of synchronization phenomena in such systems is of high practical importance. Time delay systems are also interesting because the dimension of their chaotic dynamics can be made arbitrarily large by increasing the delay time [4]. From this point of view these systems are especially appealing for secure communication schemes [5].

The recently discovered lag synchronization [6], which appears as a coincidence of shifted-in-time states of two systems, $y(t) = x_\tau \equiv x(t - \tau)$ with positive $\tau$ between two bi-directionally coupled chaotic systems described by ordinary differential equations with parameter mismatch is *approximate* in nature, as assumptions concerning averaging over rotations of the phases and constant slow phase assumption allowed the authors of [6] to obtain this type of synchronization. Although in recent years lag synchronization in time delay systems has been studied intensively both experimentally and numerically in representative systems such as chaotic semiconductor lasers with optical feedback [5], no analytical treatment of *exact* (i.e. without any approximations) lag synchronization has been performed. Most experimental investigations on chaos synchronization in unidirectionally coupled an external cavity semiconductor lasers [5] have found that the lag time between the master and slave lasers is equal to the coupling delay, whereas numerical results [7] show that the lag time should be equal to the difference between the delay in the coupling and round-trip time of the light in the transmitter's external cavity for the coupled identical systems.

---

[1] Permanent address: Institute of Physics, 370143 Baku, Azerbaijan
[2] Electronic address: alan@sees.bangor.ac.uk

Knowledge of the exact lag time is of considerable practical importance, as experiments on message transmission using fibre lasers and diode lasers have shown that the recovery of message at the receiver critically depends on the correction made for the lag time [5,8]. In a recent paper [9] *numerical* study of two distant unidirectionally coupled single-mode semiconductor lasers subject to optical feedback was reported. In [9] it was found that two fundamentally different types of chaotic lag synchronization can occur depending on the strengths of the coupling and of the feedback of the receiver laser. In the first type of synchronization, when the feedback rates of the transmitter and receiver lasers are equal, the lag time is equal to the coupling delay between the transmitter and receiver lasers; in the second type of synchronization, when the feedback rate of the transmitter is equal to the sum of the feedback rate of the receiver and coupling strength, the lag time is the difference between the coupling delay and the round-trip time of the light in the transmitter. The first type of lag synchronization was studied experimentally in [5]. The second type of lag synchronization was observed experimentally in [10]. However as it follows from the results of numerical simulations of the first type of lag synchronization the synchronization error approaches zero only occasionally even after introducing a constant multiplying correction coefficient. Thus as the authors of [9] themselves acknowledge the synchronization manifold introduced in [9] is not exact.

In this Letter we develop *an analytical approach* to *exact* lag synchronization in unidirectionally coupled time delay systems governed by two characteristic delay times, where the delay time in the coupling is different from the delay time in the coupled systems themselves. We find that for identical systems (no parameter mismatches) the lag time is equal to the difference between the coupling delay time and the time delay in the coupled systems themselves. We pay particular attention to the investigation of lag synchronization in chaotic external cavity semiconductor lasers and for the first time demonstrate that parameter mismatches can explain why the experimentally measured lag time between the transmitter and the receiver waveforms is equal to the coupling delay and is independent of the round-trip time of the light in the external cavity of the transmitter.

To enhance the accessibility and practicality of our presentation, we demonstrate the principles using concrete examples from different areas of physics. First consider the following unidirectionally coupled driver (x) and response (y) systems with different feedback and coupling delays $\tau_1$ and $\tau_2$, respectively [11].

$$\frac{dx}{dt} = -\alpha x + m_1 \sin x_{\tau_1},$$
$$\frac{dy}{dt} = -\alpha y + m_2 \sin y_{\tau_1} + m_3 \sin x_{\tau_2}, \qquad (1)$$

where $m_1, m_2$ and $m_3$ are constants; $\alpha > 0$. Equations (1) play an important role in electronics and physiological studies [12]. Now we demonstrate that for $\tau_2 > \tau_1$

$$y = x_{\tau_2 - \tau_1} \qquad (2)$$

is the lag synchronization manifold. From eqs.(1) it follows that under conditions

$$m_1 = m_2 + m_3 \tag{3}$$

the error $\Delta = x_{\tau_2-\tau_1} - y$ satisfies the following time delay equation

$$\frac{d\Delta}{dt} = -\alpha\Delta + m_2 \cos x_{\tau_1} \Delta_{\tau_1}. \tag{4}$$

To study the stability of the lag synchronization manifold one can use a Krasovskii-Lyapunov functional approach. According to [4], the sufficient stability condition for the trivial solution $\Delta = 0$ of time delay equation $\frac{d\Delta}{dt} = -r(t)\Delta + s(t)\Delta_\tau$ is: $r(t) > |s(t)|$. Thus we obtain that

$$\alpha > |m_2| \tag{5}$$

is the sufficient stability condition for lag synchronization and the lag synchronization manifold $y = x_{\tau_2-\tau_1}$ is asymptotically stable. The condition (3) is the existence (necessary) condition of lag synchronization between the unidirectionally coupled modified Ikeda model (1). Notice that in contrast to [6] we make no approximations to obtain lag synchronization.

In the following example we demonstrate lag synchronization in non-chaotic time delay systems. Consider the following unidirectionally coupled non-chaotic driver (x) and response (y) systems:

$$\frac{dx}{dt} = -\alpha_1 x + k_1 x_{\tau_1},$$

$$\frac{dy}{dt} = -\alpha_2 y + k_2 y_{\tau_1} + k_3 x_{\tau_2}, \tag{6}$$

where $k_1$ and $k_2$ are feedback rates for the driver and response systems, respectively; $k_3$ is the coupling rate. We argue that for $\alpha_1 = \alpha_2 = \alpha$, $x_{\tau_2-\tau_1} = y$ is the lag synchronization manifold. The proof comes from the investigation of the dynamics of the error $\Delta = x_{\tau_2-\tau_1} - y$ written as $\frac{d\Delta}{dt} = -\alpha\Delta + k_2\Delta_{\tau_1}$ under the existence condition $k_1 = k_2 + k_3$. Again the Krasovskii-Lyapunov functional approach allows to find the sufficent stability condition for the lag synchronization manifold: $\alpha > |k_2|$.

Next we consider lag synchronization between chaotic semiconductor lasers with optical feedback. Studying synchronization in chaotic semiconductor lasers is of great practical importance, due to their applications in high-speed optical communications and potential in chaos-based secure communications [5]. External cavity laser diodes are commonly modeled with the Lang-Kobayashi equations, see, e.g. [11]. Based on [13] here we will use rate equations for the laser intensity $I$ and excess carrier density $n$ to describe semiconductor lasers with optical feedback. The use of the rate equations, which neglect the optical phase is justified in detail in [13]. (We have also investigated the lag chaos synchronization regime in external cavity lasers modelled by the Lang-Kobayashi equations, which includes the optical phase and have derived the same existence condition for the lag synchronization manifold as in the case of the simpler rate equations presented here.) Suppose that the master laser

$$\frac{dI_1}{dt} = (gn_1 - \gamma)I_1 + k_1 I_1(t - \tau_1),$$

$$\frac{dn_1}{dt} = a - \gamma_e n_1 - gn_1 I_1 \qquad (7)$$

is coupled unidirectionally to the slave laser

$$\frac{dI_2}{dt} = (gn_2 - \gamma)I_2 + k_2 I_2(t - \tau_1) + k_3 I_1(t - \tau_2),$$

$$\frac{dn_2}{dt} = a - \gamma_e n_2 - gn_2 I_2, \qquad (8)$$

where $g$ is the diffential optical gain; $\tau_1$ is the master and slave lasers' external cavity round-trip time; $\tau_2$ is the coupling time between lasers; $\gamma_e$- the carrier density rate; $\gamma$-the cavity decay rate(cavity losses); $a$-the injection current related term; $k_1$, $k_2$ and $k_3$ are feedback and coupling rates, respectively.

Now we analytically prove that $I_2 = I_{1,\tau_2-\tau_1}$, $n_2 = n_{1,\tau_2-\tau_1}$ is the lag ($\tau_1 < \tau_2$) synchronization manifold for the interacting laser systems (7) and (8) and derive the necessary condition for that. We define the intensity and carrier errors by $\Delta_1 = I_2 - I_{1,\tau_2-\tau_1}$ and $\Delta_2 = n_2 - n_{1,\tau_2-\tau_1}$. Then under the condition

$$k_1 = k_2 + k_3 \qquad (9)$$

it is possible to obtain the following error dynamics:

$$\frac{d\Delta_1}{dt} = -\gamma \Delta_1 + gn_1 \Delta_1 + gI_1 \Delta_2 - g\Delta_1 \Delta_2 + k_2 \Delta_{1,\tau_1},$$

$$\frac{d\Delta_2}{dt} = -\gamma_e \Delta_2 - gn_1 \Delta_1 - gI_1 \Delta_2 + g\Delta_1 \Delta_2, \qquad (10)$$

where $n_1$ and $I_1$ are the solutions of eqs. (7). It is obvious that $\Delta_1 = \Delta_2 = 0$ is the solution of the system (10). We underline that while deriving (10) we did not assume that the $\Delta_1$ and $\Delta_2$ are small. The condition (9) is the necessary condition for synchronization between unidirectionally coupled identical master and slave laser systems and this analytical result is in excellent agreement with numerical simulations in [7,9] and experiments in [10]. Having found the necessary condition (existence condition) for lag synchronization, one should study the question of stability of the synchronization manifold. We add the error equations (10) to arrive at the decoupled error dynamics

$$\frac{d\Delta_1}{dt} + \gamma \Delta_1 - k_2 \Delta_{1,\tau_1} = -\frac{d\Delta_2}{dt} - \gamma_e \Delta_2. \qquad (11)$$

Equation (11) has a simple geometrical interpretation which is very helpful to find the stability condition of the inverse anticipating synchronization manifold. The trivial solutions $\Delta_1 = \Delta_2 = 0$ are the intersection points of a family of curves $\frac{d\Delta_1}{dt} + \gamma \Delta_1 - k_2 \Delta_{1,\tau_1} = c_1$ and $-\frac{d\Delta_2}{dt} - \gamma_e \Delta_2 = c_2$. As physically $\gamma_e$ is positive, we effectively arrive at the investigation of the stability of the trivial solution of equation $\frac{d\Delta_1}{dt} + \gamma \Delta_1 - k_2 \Delta_{1,\tau_1} = 0$. A Krasovskii-Lyapunov functional approach gives the sufficient stability condition: $\gamma > |k_2|$.

Now we consider lag synchronization in cascaded systems. It can be envisaged that in an advanced communication system using chaotic oscillators there will be a need to broadcast a message to a number of receivers or else use may be made of repeater stations to extend the transmission distance. In both these cases there is a need to demonstrate a capability for transferring synchronization from a master system to a number of slave systems. Here we demonstrate that synchronization can be effected between master and slave systems arranged in series. This we term cascaded synchronization. Such a arrangement would be applicable in a chain of repeaters. Let the driven system $y$ from eqs.(1) drive another response system $z$:

$$\frac{dz}{dt} = -\alpha z + m_4 \sin z_{\tau_1} + m_5 \sin y_{\tau_2}. \tag{12}$$

Then by making use of error dynamics approach above one can obtain the necessary and sufficient stability conditions for lag synchronization between the master system $x$ and the response systems $z$, $z = x_{2(\tau_2 - \tau_1)}$ with lag time $2(\tau_2 - \tau_1)$: $m_1 = m_4 + m_5$ and $\alpha > |m_4|$, respectively. Here we have assumed that lag synchronization between $x$ and $y$ state variables has already taken place; then using $y = x_{\tau_2 - \tau_1}$ we have replaced $x_{2\tau_2 - \tau_1}$ by $y_{\tau_2}$. Under this assumption it is also easy to check that the synchronization between the response systems $y$ and $z$ occurs with the lag time $\tau_2 - \tau_1$: $z = y_{\tau_2 - \tau_1}$. The existence and sufficient stability conditions for the synchronization between the response systems are: $m_2 + m_3 = m_4 + m_5$ and $\alpha > |m_4|$, respectively. It is straightforward to check that the concept of cascaded lag synchronization can be also applied to the chain of chaotic laser systems,i.e. by adding another slave laser system of the form $\frac{dI_3}{dt} = (gn_3 - \gamma)I_3 + k_4 I_3(t - \tau_1) + k_5 I_1(t - \tau_2); \frac{dn_3}{dt} = a - \gamma_e n_3 - g n_3 I_3$ to eqs.(7) and (8) one can obtain the lag synchronization manifold $I_3 = I_{1,2(\tau_2 - \tau_1)}$, $n_3 = n_{1,2(\tau_2 - \tau_1)}$ with the lag time $2(\tau_2 - \tau_1)$.). Cascaded lag synchronization of chaotic external cavity laser diodes has been recently demonstrated experimentally [14].

Thus we demonstrate that in the case of lag synchronization between two unidirectionally coupled identical time delay systems with two characteristic delay times the lag time is the difference between the coupling delay time and the delay time in the coupled systems themselves. However, most experimental investigations on chaos synchronization in unidirectionally coupled external cavity semiconductor lasers [5] have found that the lag time between the master and slave lasers is equal to the coupling delay and is independent of the round-trip time of the light in the external cavity of the transmitter. Despite active research in the field, this outstanding problem thus far remains unresolved. This problem is also outstanding for the recently reported first experimental observation of anticipating chaos synchronization in semiconductor lasers with optical feedback

[15].

In this Letter for the first time we demonstrate that parameter mismatches can explain the experimental observation that the lag time is equal to the coupling delay. First we demonstrate the idea for the example of coupled non-chaotic systems (6). Suppose that the parameter $\alpha$ is different for nonchaotic driver (master) and response (slave) systems: $\alpha_1 = \alpha - \delta$ and $\alpha_2 = \alpha + \delta$. Above we have obtained that with identical $\alpha_1$ and $\alpha_2$ the system (8) allows for the lag synchronization manifold $y = x_{\tau_2-\tau_1}$ with the lag time $\tau_2 - \tau_1$. Now we argue that with parameter mismatches i.e. $\alpha_1 \neq \alpha_2$ system (6) allows for the lag synchronization manifold $y = x_{\tau_2}$ with the lag time $\tau_2$. Denote the error signal by $\Delta = x_{\tau_2} - y$. Then from eq.(6) one can obtain that

$$\frac{d\Delta}{dt} = -(\alpha + \delta)\Delta + 2\delta x_{\tau_2} + k_1 x_{\tau_1+\tau_2} - k_2 y_{\tau_1} - k_3 x_{\tau_2}. \tag{13}$$

Thus under the existence conditions $k_1 = k_2$ and $2\delta = k_3$ for lag synchronization manifold $y = x_{\tau_2}$ from the error $x_{\tau_2} - y$ dynamics

$$\frac{d\Delta}{dt} = -(\alpha + \delta)\Delta + k_1 \Delta_{\tau_1}. \tag{14}$$

we also obtain the sufficient stability condition: $\alpha + \delta = \alpha_2 > |k_1|$. On this basis we now consider the systems (7) and (8). We suppose that the cavity decay rate $\gamma$ in eqs.(7) and (8) is different for the driver ($\gamma_1$) and response ($\gamma_2$) systems: $\gamma_1 = \gamma - \delta$ and $\gamma_2 = \gamma + \delta$, where $\delta$ determines the parameter mismatch (usually different values for the cavity decay rates are used by different authors, see e.g.[7,11]). Now we argue that with parameter mismatches, $I_2 = I_{1,\tau_2}$ can be the lag synchronization manifold for systems (7) and (8) with the lag time being equal to the coupling delay. From eqs.(7) and (8) it is possible to obtain, in analogy to eq.(11), decoupled error dynamics for $\Delta_1 = I_2 - I_{1,\tau_2}$ and $\Delta_2 = n_2 - n_{1,\tau_2}$:

$$\frac{d\Delta_1}{dt} + (\gamma+\delta)\Delta_1 - k_1 I_{1,\tau_1+\tau_2} + k_2 I_{2,\tau_1} - (2\delta - k_3)I_{1,\tau_2} = -\frac{d\Delta_2}{dt} - \gamma_e \Delta_2. \tag{15}$$

In other words $\Delta_1 = \Delta_2 = 0$ is the solution of eq.(15) if the existence conditions $k_1 = k_2$ (under this condition $k_2 I_{2,\tau_1} - k_1 I_{1,\tau_1+\tau_2} = k_2 \Delta_{1,\tau_1}$) and $2\delta = k_3$ are satisfied. Thus we obtain that the lag time between the waveforms in the master and slave lasers is equal to the coupling delay, if the parameter mismatch $\gamma_2 - \gamma_1$ equals the coupling strength $k_3$: i.e. $2\delta = k_3$ and the feedback rates for the master and slave lasers are equal $k_1 = k_2$. This analytical result is in very good agreement with numerical simulations in [9] and experiments in [5]. As one can easily estimate the coupling rate $k_3$ in the experiments the condition $2\delta = k_3$ can also be used for the estimation of the differences of the cavity losses between the semiconductor lasers to be synchronized.

The following point deserves to be underlined. In [9] it is indicated that numerical simulations of the second type of synchronization contrasts with Masoller's numerical work on anticipating synchronization [11], the phenomenon discovered by Voss in [12]. According to [9], Masoller has pointed out that the receiver anticipates the electric field produced by the transmitter only if $\tau_2 < \tau_1$. (It can be proven analytically that under the condition (9) $I_1 = I_{2,\tau_1-\tau_2} \equiv I_2(t-(\tau_1-\tau_2))$, $n_1 = n_{2,\tau_1-\tau_2}$ is the anticipating ($\tau_1 > \tau_2$) synchronization manifold for the interacting laser systems (7) and (8).) In [9] it is argued that whatever the value of $\tau_2$, the output of the receiver $I_2$ anticipates the signal that is injected into it $I_{1,\tau}$ with an anticipation time $\tau_1$. Indeed it can be checked that under the condition (9), which coincides with the condition (8) in [9] that $I_{1,\tau_2} = I_{2,\tau_1}$, $n_{1,\tau_2} = n_{2,\tau_1}$ is the synchronization manifold for the systems (7) and (8), whatever the value of $\tau_2$. In fact there is no contradiction between Masoller's numerical results and the numerical simulations in [9], as the manifolds $I_1 = I_{2,\tau_1-\tau_2}$ and $I_{1,\tau_2} = I_{2,\tau_1}$ are equivalent.

Finally, applying the procedure above to the coupled chaotic Ikeda systems [12]

$$\frac{dx}{dt} = -\alpha_1 x - \beta \sin x_{\tau_1},$$
$$\frac{dy}{dt} = -\alpha_2 y - \beta \sin y_{\tau_1} + K x_{\tau_2}, \tag{16}$$

one can obtain that $y = x_{\tau_2}$ is the lag synchronization manifold, if the parameter mismatch $\alpha_2 - \alpha_1 = 2\delta$ is equal to the coupling rate $K$. (Notice that in this example without the parameter mismatch, i.e. $\alpha_1 = \alpha_2 = \alpha$ neither $y = x_{\tau_2-\tau_1}$ nor $y = x_{\tau_2}$ is the synchronization manifold.) This can be seen by the error dynamics:

$$\frac{d\Delta}{dt} = -(\alpha+\delta)\Delta + (2\delta - K)x_{\tau_2} - \beta \cos(x_{\tau_1+\tau_2})\Delta_{\tau_1}. \tag{17}$$

The sufficient stability condition for the lag synchronization manifold $y = x_{\tau_2}$ can be written as: $\alpha + \delta = \alpha_2 > |\beta|$.

To summarize, we have presented an analytical investigation of *exact* lag synchronization between unidirectionally coupled time delay systems with two characteristic delay times, where the delay time in the coupling is different from the delay time in the coupled systems themselves. We have shown that the lag time between the oscillations of the master and slave systems is the difference between the coupling delay time and the delay time in the coupled systems themselves. We have also considered lag chaotic synchronization in cascaded systems and derived the existence and sufficient conditions for such possible practical communication schemes. Also, for the first time we have demonstrated that parameter mismatches can explain the experimental observation that the lag time is equal to the coupling delay.

This work is supported by UK EPSRC under grants GR/R22568/01 and GR/N63093/01.